# Evolution from sinusoidal to collinear A-type antiferromagnetic spin-ordered magnetic phase transition in $Tb_{0.6}Pr_{0.4}MnO_3$

**Harshit Agarwal[1†], José Antonio Alonso[2], Ángel Muñoz[3], R J Choudhary[4], O N Srivastava[1], M A Shaz[1*]**

[1]*Department of Physics, Institute of Science, Banaras Hindu University, Varanasi, 221005, India*

[2]*Instituto de Ciencia de Materiales de Madrid, CSIC, Cantoblanco, E-28049, Madrid, Spain*

[3]*Departamento de Física Aplicada, EPS, Universidad Carlos III, Avenida Universidad 30, E-28911, Leganés-Madrid, Spain*

[4]*UGC-DAE Consortium of Scientific Research, Indore, 452017, India*

## Abstract:

The present study reports on the structural and magnetic phase transitions in Pr-doped polycrystalline $Tb_{0.6}Pr_{0.4}MnO_3$, using high-resolution neutron powder diffraction (NPD) collected at SINQ spallation source (PSI), to emphasize the suppression of the sinusoidal magnetic structure of pure $TbMnO_3$ and the evolution to a collinear A-type antiferromagnetic ordering. The phase purity, Jahn-Teller distortion, and one-electron bandwidth for $e_g$ orbital of $Mn^{3+}$ cation have been calculated for polycrystalline $Tb_{0.6}Pr_{0.4}MnO_3$, in comparison to the parent materials $TbMnO_3$ and $PrMnO_3$, through the Rietveld refinement study from X-ray diffraction data at room temperature. The temperature-dependent zero field-cooled and field-cooled *dc* magnetization study at low temperature down to 5 K reveals a variation in the magnetic phase transition due to the effect of $Pr^{3+}$ substitution at the $Tb^{3+}$ site, which gives the signature of the antiferromagnetic nature of the sample, with a weak ferromagnetic component at low temperature induced by an external magnetic field. The field-dependent magnetization study at low temperatures gives the weak coercivity having the order of 2 kOe, which is expected due to canted-spin arrangement or ferromagnetic nature of Terbium ordering. The NPD data for $Tb_{0.6}Pr_{0.4}MnO_3$ confirms that the nuclear structure of the synthesized sample maintains its orthorhombic symmetry down to 1.5 K. Also, the magnetic structures have been solved at 50 K, 25 K, and 1.5 K through the NPD study, which shows A-type antiferromagnetic spin arrangement.

**Email:** [*]shaz2001in@yahoo.com

[†]harshit.physics@gmail.com

## 1. Introduction:

The emerging trend of multiferroic materials develops its specific role in material science research due to the coexistence of a spatially modulated magnetic order and a uniform polarization induced by broken inversion symmetry [1,2]. Perovskite-structured-based rare-earth manganites are a promising case within the multiferroic family because, not only their multiferroic properties are quite prominent, but also the involved physical mechanisms are very interesting in this field of research [3,4]. These $RMnO_3$ manganites show multiferroicity usually at low temperatures and can be used in various applications of spintronics [5], magnetic sensing, and storage devices [2,3,6,7]. The magnetic ordering in the rare-earth manganites depends upon the cationic size of $R^{3+}$ ion. Also, the $Mn^{3+}$ ordering has a strong competition in between the AFM coupling due to $t_{2g}$ electrons and the FM coupling due to $e_g$ electrons[8]. Such AFM-FM competition and varying size of $R^{3+}$ ion transform the magnetic structure of rare-earth manganites, changing from layered A-type antiferromagnetic to incommensurate spiral magnetic structure and then switches into E-type AFM magnetic structure for higher-Z rare-earth ions[9]. For example, $PrMnO_3$ exhibits an A-type antiferromagnetic structure at $T_N$ = 100 K in which the magnetic spins are aligned in ferromagnetic order along the x-z plane and antiferromagnetically ordered along the y-axis[10]. Similarly, a structural transition takes place in $NdMnO_3$ at $T_N$= 80 K [11]. In more distorted compounds such as R= Dy, Tb, and Gd, the magnetic structure becomes sinusoidally modulated in the a-b plane below 39-43 K, becoming spirally modulated below 18-17 K [9,12]. $TbMnO_3$ is the prominent example of type II spiral multiferroic, which shows large magneto-electric coupling as the ferroelectricity in terbium manganite develops due to the application of a spiral magnetic spin ordering, which breaks the spatial inversion symmetry [13]. In $TbMnO_3$, the magnetic order is associated with an incommensurate wave vector $k_{Mn}$ = (0, q, 0) with q = 0.27, showing a sinusoidal modulated magnetic structure due to $Mn^{3+}$ ordering at around 25 K. Another wave vector $k_{Tb}$ = (0, $\tau$, 0) with $\tau$ = 0.425 for $Tb^{3+}$ ordering is concomitant to the canted spiral magnetic structure due to inverse Dzyaloshinskii-Moriya (DM) coupling [14]. The structural correlation to magneto-electric coupling in polycrystalline $TbMnO_3$ has been discussed using low-temperature high magnetic field X-ray diffraction as well as a magneto-dielectric study [15].

Several studies on neutron diffraction experiments correlated to physical properties have been reported for dopant-mediated rare-earth manganites[16]. It was observed that the ionic radii and Mn-$O_6$ octahedral distortion due to the variation of $Mn^{3+}$ - O - $Mn^{3+}$ super-exchange interactions are responsible for the structural and magnetic parameters of these $RMnO_3$ manganites [17]. The occurrence of mixed $RE^{3+}$ rare-earth ions in $RMnO_3$ affects the magnetic ordering of $R^{3+}$ ions and indirectly modifies the $Mn^{3+}$ ordering [18]. A structural and magnetic phase transition was observed in Pr-doped $LaMnO_3$, where a low-temperature disordered system leads to a spin canted feature, as a result of the variance in A site cationic radii [19]. Several studies have been undertaken on $Dy^{3+}$-doped $TbMnO_3$ related to collinear to a spiral-spin magnetic phase transition around the ferroelectric transition [20], exchange bias effect [21], and magneto-caloric study [22]. Recently, the effect of magnetization spin reversal has been observed through a neutron diffraction study, observed in hole-doped polycrystalline $TbMnO_3$ [23]. There is a lack of understanding of the role of lower Z rare-earth ions (as- $Pr^{3+}$, $Nd^{3+}$) in systematic studies of doping at the $Tb^{3+}$ site in $TbMnO_3$. However, the study of $Nd^{3+}$ doped $TbMnO_3$ reveals the incommensurate magneto-electric phase of $TbMnO_3$ and the suppression of the magnetic ordering by substituting the $Nd^{3+}$ at $Tb^{3+}$ site [24]. The present study reports on the role of $Pr^{3+}$ doping at the $Tb^{3+}$ site of $TbMnO_3$ in changing the magnetic phase transition at low temperature, which has not been reported before. We have used high-resolution neutron powder diffraction (NPD) data for understanding the nuclear and magnetic structure at low temperatures, in complement with a field- and temperature-dependent DC magnetization investigation.

## 2. Experimental:

We have substituted the trivalent cation $Pr^{3+}$ at the $Tb^{3+}$ site of $Tb_{1-x}Pr_xMnO_3$ for x= 0.4. Single-phase $Tb_{0.6}Pr_{0.4}MnO_3$ oxide, as well as the parent compounds i.e. $TbMnO_3$ and $PrMnO_3$, have been synthesized by the solid-state synthesis method, using high-purity oxide precursors (≥ 99.9% or better), $Tb_4O_7$, $MnO_2$, $Pr_2O_3$. The oxide precursors were preheated at 150° C for 24 hours, then mixed in proper stoichiometry in an agate mortar and pestle; the mixture was ground for 8 hours. Then the mixture was placed in a high-temperature furnace by 'Metrex' and calcined at 1300 ˚C for 24 h. For homogeneity and single-phase formation, the calcined sample was ground again for 2 hours and pelletized by applying 6-ton pressure for 20 min. The pellet was sintered at 1400 ˚C for 24 h, followed by furnace cooling. The mentioned parent materials have also been synthesized by solid-state synthesis from the corresponding oxide precursors, using the same procedure. The synthesized samples of pure $TbMnO_3$ (TMO), Pr doped $Tb_{0.6}Pr_{0.4}MnO_3$ (TPMO) and $PrMnO_3$ (PMO) have been characterized by X-ray diffraction (XRD) to confirm the crystallinity and phase purity, in a Panalytical Empyrean X-ray diffractometer with a step size of 0.02˚ 2θ in the 2θ range from 10˚ - 110˚. The XRD patterns of the synthesized samples were further analyzed by the Rietveld refinement technique to evaluate the lattice and structural distortion of TPMO in comparison to TMO and PMO. The further study is focused on TPMO. For the magnetization study, the temperature-dependent DC magnetization was carried out in the temperature range 5-300 K at 100 Oe, 5000 Oe, 10 kOe using a quantum design SQUID VSM (70 kOe). Also, the DC Magnetization study of polycrystalline TPMO sample has been measured as a function of applied magnetic field M(H) (-70 kOe to 70 kOe at different temperatures) by SQUID-VSM (M/s. Quantum design, USA).

We have carried out temperature-dependent neutron diffraction on polycrystalline $Tb_{0.6}Pr_{0.4}MnO_3$ at the High-Resolution Powder Diffractometer for Thermal Neutrons (HRPT) in the SINQ spallation source at PSI, Switzerland. For understanding the magnetic structure of $Tb_{0.6}Pr_{0.4}MnO_3$ at low temperature through neutron scattering, we have filled around 4 g of sample in a vanadium can with a collection time of 3 h required for each temperature, ranging from 300 K down to 1.5 K. The wavelength of the incident neutron beam was 1.494 Å. The analysis of the nuclear and magnetic refinement was performed by using the program Fullprof [25,26]. The magnetic propagation vector for analyzing the magnetic reflections has been determined by K-search from the peak positions of the magnetic diffraction lines, and the corresponding irreducible representations were determined by BASIREPS [27] modules in Fullprof crystallographic suite.

## 3. Result and discussion

### 3.1 Structural observation using X-ray diffraction

A comparison in the X-ray diffraction patterns of $TbMnO_3$ (TMO), Pr doped $Tb_{0.6}Pr_{0.4}MnO_3$ (TPMO), and $PrMnO_3$ (PMO) are shown in Figure 1 (a). The inset highlights the shift of Bragg peaks to a lower angle from TMO to doped TPMO and PMO. All the XRD patterns have been refined using the Rietveld refinement method and found that the materials are synthesized as single phasic materials, crystallizing in an orthorhombic structure with centrosymmetric *Pnma* symmetry. The Rietveld refinement of TPMO is displayed in Figure 1 (b). All peaks are well fitted with no impurity peaks present in the XRD pattern of polycrystalline TPMO (Refined Parameters: GOF= 1.09, $R_p$= 2.23%, $R_{wp}$= 2.99%). The refined structure of TPMO is shown in Figure 1 (b), where the O1 and O2 atoms occupy the two apical positions of the $MnO_6$ octahedra and the four equatorial positions, respectively. **Table 1** displays the refined structural and lattice parameters for all the samples.

The structural and physical properties of rare-earth manganites are influenced by various crystallographic factors, such as the tolerance factor, the Jahn-Teller distortion of $Mn^{3+}$ ions, the average ionic radii of cations, and fractional occupancies[28,29]. The tolerance factor (t) based on the average ionic radii of $\langle r_A \rangle$, $\langle r_B \rangle$ and $\langle r_O \rangle$ has been calculated for TMO, TPMO and PMO as the tolerance factor; t = 0.863, 0.874, and 0.892, respectively, which allows evaluating how these structures are distorted from cubic symmetry (t = 1) to orthorhombic symmetry. Also, the orthorhombic strain of all three samples can be calculated by the spontaneous strain parameter based on the lattice parameters, which evaluates the strain induced in the orthorhombic structure of the concerned material. In the case of *Pnma* space group, the orthorhombic strain is s = [(a – c) / (a + c)], taking the values s = 0.097, 0.082 and 0.0104 for TMO, TPMO and PMO, respectively. This means that the orthorhombic structure of TMO is more severely distorted than TPMO and PMO. **Figures 1 (c), (d), and (e)** are showing the top view of the octahedral tilting in the crystal structure of TMO, TPMO, and PMO, having the tilting scheme of $a^-b^+a^-$ as per the Glazer's notations[30]. The octahedral tilting is decreasing from TMO to TPMO and in PMO, which suggests that the Mn-O octahedral distortion in TPMO lies in between the distortion of TMO and PMO. **Table 2** displays the geometrical parameters characterizing the crystal structure of TMO, TPMO, and PMO at room temperature. We have found that, by increasing the average ionic radius $\langle r_A \rangle$ of the A-site ion, the tilting of Mn-$O_6$ octahedra decreases, and the Mn-O-Mn bond angle increases. The octahedral tilting combined with the static Jahn-Teller (JT) distortion produced by super-exchange interaction of $Mn^{3+}$-O-$Mn^{3+}$, can be calculated using the relation[31],

$$\Delta_{JT} = \frac{1}{N} \sum_{i=1 \ldots N} \left[ \frac{(Mn-O)_i - \langle Mn-O \rangle}{\langle Mn-O \rangle} \right]^2$$

The JT distortion is decreasing hereby increasing the cationic radius of A site, as shown in Figures 1 (c-e), which results in the variation of the one-electron bandwidth of $e_g$ band of $Mn^{3+}$ ion. The one-electron bandwidth of $e_g$ band can be calculated by[32],

$$W = \frac{\cos(\omega/2)}{d^{3.5}_{<Mn-O>}} \text{ , where } \omega = \pi - <Mn - O - Mn>$$

The one-electron bandwidth W for TPMO lies in between the values for TMO and PMO as observed in **Table 2**. The lower bandwidth favors the antiferromagnetic insulating state by localizing the $e_g$ charge carriers. The phase diagram of the thermal variation of $RMnO_3$ as a function of the Mn-O-Mn bond angle shows how the Mn-O-Mn angle is affecting the spin state of the $RMnO_3$ oxides [33,34]. The magnetic structure of $TbMnO_3$ is sinusoidal or incommensurate spiral antiferromagnetic and the structure of $PrMnO_3$ is A-type antiferromagnetic. It is clear from the Mn-O-Mn bond angle and JT distortion observation of Pr substituted TPMO that it is placed in between the values of TMO and PMO. So, we may expect that the magnetic structure of TPMO must have spin ordering in the mean state of TMO and PMO, having the magnetic transition below 100 K.

### 3.2 *dc* Magnetization study of $Tb_{0.6}Pr_{0.4}MnO_3$

For the study of the magnetic phase transition in polycrystalline TPMO, the temperature-dependent DC magnetization measurements have been collected at various fields of 100 Oe, 5000 Oe, and 10 kOe. Generally, in the case of rare-earth manganites, as the materials are cooled down from room temperature to lower temperatures, the $Mn^{3+}$ sublattice ordering is observed at $T_N$, which decreases monotonically from 145 K in $LaMnO_3$ to 45 K in $HoMnO_3$, whereas the $R^{3+}$ sublattice orders at a much lower temperature, generally below 20K [33,35]. As we have substituted $Pr^{3+}$ at the site of $Tb^{3+}$, the Mn ordering temperature must lie in this temperature range. **Figure 2** shows the thermal variation of the magnetization in the sample at zero fields cooled and field cooled state in the temperature range of 300 K – 5 K, measured under an external field of 100 Oe. The sample was cooled down to 5 K and a magnetic field of 100 Oe has been applied to the system. The zero-field cooled (ZFC) data were measured from 5 K to 300 K, after which the system was cooled down again with an applied magnetic field and the field-cooled (FC) magnetization data was recorded in the warming run. The ZFC and FC curves start to diverge around 70 K, after which the magnetic ordering start to develop in the material, as shown in figure 2 (a). The first derivative of the ZFC magnetization curve is shown in Figure 2 (b), which is used for clarifying the transition temperatures present in TPMO. The inverse susceptibility of TPMO is shown in Figure 2 (c), which exhibits the antiferromagnetic nature of the material, as observed from the Curie Weiss fitting. The Curie-Weiss fitting gives a negative Curie-Weiss temperature, which is the signature of the AFM nature of the sample. The theoretical paramagnetic effective magnetic moment $\mu_{eff}$ for TPMO can be calculated as $\sqrt{(0.6\mu_{Tb^{3+}}^2) + (0.4\mu_{Pr^{3+}}^2) + (0.6\mu_{Mn^{3+}}^2) + (0.4\mu_{Mn^{3+}}^2)}$ where $<\mu_{Tb}{}^{3+}> = 9.72\ \mu_B$, $<\mu_{Pr}{}^{3+}> = 3.58\ \mu_B$ and $<\mu_{Mn}{}^{3+}> = 4.89\ \mu_B$. The experimentally calculated $\mu_{eff}$ for TPMO is 9.67(6) $\mu_B$, which is close to the theoretical value $\mu_{eff}$ for TPMO, of 9.25(8) $\mu_B$. In the inset of Figure 2 (c) a clear intrinsic magnetic behavior of TPMO is shown at low temperatures. There are multiple magnetic transitions are observed in the regions i, ii, iii, iv, and v of inverse susceptibility observed ZFC state of 100 Oe, which indicates the material is approaching from antiferromagnetic to ferromagnetic spin state by lowering the temperature as observed by the Curie-Weiss fitting of these regions. Figure 2 (d) represents the difference between the ZFC and FC curves of the material. There is a large difference between ZFC and FC magnetization of TPMO, which is the characteristic of magnetic relaxation produced in the sample. The separation between ZFC and FC magnetization is increasing down to 5 K, which indicates an enhancement of the ferromagnetic content in the material at lower temperatures, in a matrix dominated by antiferromagnetic interactions.

The material turns from a paramagnetic state to an AFM state due to Mn ordering at $T_N$ = 50 K. When we decrease the temperature, we have found that there are other transitions, observed at around $T_{Cusp}$ = 32 K, where the AFM transition enhanced due to Mn-Mn interaction; and $T_{Tb}$= 13 K, where a ferromagnetic component appears associated with the ordering of the $Tb^{3+}$ ions (as visible in the thermal variation of the first derivative of magnetization in the ZFC state). Such ferromagnetic contribution is expected to further enhance down to 1.5 K.

When the DC magnetization study is carried out at a higher magnetic field viz 5000 Oe and 10 kOe, the transition temperature that we have observed at 100 Oe seems to shift towards lower temperature. The thermal variation of DC magnetization and inverse susceptibility are displayed in Figure 3 at the magnetic fields of 100 Oe, 5000 Oe, and 10 kOe, in the temperature range of 5

K-100 K. The inverse susceptibility curves at 5000 Oe and 10 kOe indicate that the material is attaining the purely antiferromagnetic state, as the Curie Weiss fitting for ZFC state is moving to linearity in the complete temperature range. The multiple magnetic transitions as observed in DC magnetization of 100 Oe field seem to vanish at 5000 Oe and 10 kOe. **Table 3** shows the magnetization parameters observed by the Curie-Weiss fitting of the sample in various fields.

The magnetic-field dependent magnetization study was carried out at different temperatures down to 5 K and magnetic field up to 70 kOe, as shown in **Figure 4**. It was found that at 50 K, the material starts developing the AFM type in nature. At 30 K and 5 K, there are weak coercivities of 1954 Oe and 2231 Oe, respectively, for TPMO. The M/H curve obtained at 5 K represents a weak hysteresis, which is expected a characteristic ferromagnetic component due to Tb ordering. The magnetic structure of the TPMO is further elucidated in the low-temperature NPD investigation.

### 3.3 Neutron Diffraction study of $Tb_{0.6}Pr_{0.4}MnO_3$

To further investigate the magnetic structure of TPMO, we have performed a neutron powder diffraction (NPD) analysis of the sample at different temperatures down to 1.5 K. The NPD patterns at 300 K, 150 K, 50 K, 25 K, and 1.5 K are shown in **Figure 5 (a)**. We have found that, at 50 K and below, magnetic peaks are starting to develop at the lower angle side of the NPD pattern, due to the magnetic form factor. It was found that one magnetic peak is arising at $2\theta = 11.49°$ at 50 K, which has a higher intensity in 25 K and 1.5 K NPD patterns. This magnetic reflection appears at the forbidden position of the Bragg peak (0,1,0) for the *Pnma* space group. At T = 25 K, two other magnetic peaks are clearly observed at the positions of the Bragg peaks (1,1,1) and (0,3,0). The difference curve in between the NPD patterns of 50 K, 25 K, and 1.5 K with respect to NPD pattern of 150 K shows the magnetic reflection present in the NPD pattern; shown in figure 5 (b). The propagation vectors defining the magnetic structure is found $\mathbf{q}$ = (0,0,0) for which the nuclear and magnetic cell sizes coincide. At 1.5 K, the magnetic Bragg reflections are getting more intense and pronounced, which may be due to the Terbium ordering. The Rietveld refinement of NPD patterns is illustrated in **Figure 6 (a-e).** The nuclear peaks are well fitted with the centrosymmetric *Pnma* symmetry. The lattice and structural parameters observed by the refinement of NPD patterns are included in **Tables 4 and 5**, respectively**.**

According to the spectra, the magnetic peak (0,1,0) does not change its position excepting only its intensity; this means that the structure is stable from the ordering temperature down to 1.5 K. The possible magnetic structures compatible with the symmetry of the compound have been determined following the representation analysis technique described by Bertaut [36]. For the propagation vector **q** = (0,0,0); the small group coincides with the space group *Pnma*. The possible magnetic structures for Tb and Mn ions and each one of the irreducible representations are given in **Table 6.** As the decomposition of magnetic representation for Mn and Tb atoms in case of Pnma symmetry is shown as:

$\boldsymbol{\Gamma_{Mn}} = 3\,\Gamma_1 + 3\,\Gamma_3 + 3\,\Gamma_5 + 3\,\Gamma_7$

$\boldsymbol{\Gamma_{Tb}} = 1\,\Gamma_1 + 2\,\Gamma_2 + 2\,\Gamma_3 + 1\,\Gamma_4 + 1\,\Gamma_5 + 2\,\Gamma_6 + 2\,\Gamma_7 + 1\,\Gamma_8$

At T = 50 K and 25 K, only the Mn sublattice contributes to the magnetic ordering and, after checking the different solutions, the magnetic structure is given by the basis vector ($C_x$,$F_y$,$A_z$) belonging to the $\boldsymbol{\Gamma_5}$ representation. The best solution is the (Cx,0,0) for the Mn atoms. This solution is coherent with the magnetic reflections observed in the NPD patterns, since for the coupling of the magnetic moments of the Mn atoms given by the **C** basis vector, it is only possible to observe those magnetic reflections with k = 2n+1 and h + k = 2n. However, a slight canting in the Mn moment was expected when the $m_y$ and $m_z$ component are released to refine, but it was found negligible in comparison to $m_x$ component. The good agreement between observed and calculated magnetic intensities can be appreciated in the insets of Figure- 5d and 5e. The results of the magnetic moment observed for Mn and Tb cations at 50 K, 25 K, and 1.5 K are shown in **Table 7**. The magnetic structures corresponding to all these temperatures are displayed in **Figure 7**. It can be observed that the arrangement of the $Mn^{3+}$ spins is a collinear A-type AFM magnetic structure. This magnetic ordering is similar to the magnetic ordering exhibited by $PrMnO_3$. It means that the partial substitution of the Pr cations by Tb cations leads to the destruction of the sinusoidal magnetic structure of $TbMnO_3$. The fitting of the NPD patterns acquired at T = 1.5 K indicates that the Tb cations order according the basis vector (0,$F_y$,0) also belonging to the $\Gamma_5$ irreducible representation. So, the Tb cations exhibit a ferromagnetic coupling with the magnetic moments parallel to the b-direction, in good agreement with the magnetization measurements. The results of the fitting are shown in Table 7 and Figure 5f. The magnetic structure is stable across the transition temperatures, with little evolution upon cooling, apart from the expected increase of the magnitude of the Mn moments, as shown in **Figure 7 and Table 7**. According to the $\mathbf{\Gamma_5}$ representation, the magnetic space group of the magnetic structure is *Pn´ma´*.

The magnetic phase transition in polycrystalline $Tb_{0.6}Pr_{0.4}MnO_3$ can be observed in **Figure 8** in comparison to parent materials $TbMnO_3$ and $PrMnO_3$. The phase diagram is based on the study represented by the Kimura[34]. The study is significant to observe the further physical and transport studies as Pr doping is affecting the transition temperatures of pure TMO. Thus, by substituting the $Pr^{3+}$ ion at the Tb site of $Tb_{0.6}Pr_{0.4}MnO_3$ perovskite, the sinusoidal structure of pure $TbMnO_3$ evolves to the A-type AFM structure.

## 4 Conclusions

The polycrystalline $Tb_{0.6}Pr_{0.4}MnO_3$ has been synthesized using a solid-state synthesis method; this perovskite oxide crystallizes in the orthorhombic symmetry of space group *Pnma*. The structural observation reveals that the Mn-O octahedral distortion for $Tb_{0.6}Pr_{0.4}MnO_3$ is weaker than that observed in $TbMnO_3$ and higher than that of $PrMnO_3$. The Mn-O-Mn bond angle variation and Jahn-Teller distortion in TPMO lie in between those of the parent materials, which give the signature of the mean magnetic spin ordering state of TMO and PMO. The *dc* Magnetization study reveals the development of an AFM state at lower temperatures below 50 K. The neutron powder diffraction study confirms that the nuclear structure maintains the orthorhombic symmetry from room temperature to the low temperature down to 1.5 K. The magnetic structure at 50 and 25 K is originated by the long-range ordering of Mn spins, and it possesses an A-type antiferromagnetic arrangement, while at 1.5 K the Tb moments are also contributing to the magnetic structure, showing a ferromagnetic ordering along the b axis.

## Acknowledgments

HA and MAS are thankful to Prof. Anand Chaudhary, IMS, BHU for providing the high-temperature furnace facility. The authors are grateful to Dr. S. D. Kaushik, Prof. D. Sa, and Prof. A. K. Ghosh for fruitful discussion and providing the salient suggestions on the manuscript. Author 'HA' would like to acknowledge the CSIR India for providing a senior research fellowship (09/013(0763)/2018-EMR-I). JAA thanks the Spanish MINECO for funding Project MAT2017-84496-R.

**References:**

[1] Fiebig M, Lottermoser T, Meier D and Trassin M 2016 The evolution of multiferroics *Nat. Rev. Mater.* **1**

[2] Spaldin N A and Ramesh R 2019 Advances in magnetoelectric multiferroics *Nat. Mater.* **18** 203–12

[3] Yi D, Lu N, Chen X, Shen S and Yu P 2017 Engineering magnetism at functional oxides interfaces: Manganites and beyond *J. Phys. Condens. Matter* **29**

[4] Markovich V, Wisniewski A and Szymczak H 2014 Magnetic Properties of Perovskite Manganites and Their Modifications *Handbook of Magnetic Materials* vol 22 pp 1–201

[5] Fusil S, Garcia V, Barthélémy A and Bibes M 2014 Magnetoelectric devices for spintronics *Annu. Rev. Mater. Res.* **44** 91–116

[6] Jia T, Cheng Z, Zhao H and Kimura H 2018 Domain switching in single-phase multiferroics *Appl. Phys. Rev.* **5**

[7] Zvezdin A K, Logginov A S, Meshkov G A and Pyatakov A P 2007 Multiferroics: Promising materials for microelectronics, spintronics, and sensor technique *Bull. Russ. Acad. Sci. Phys.* **71** 1561–2

[8] Aliouane N, Prokhnenko O, Feyerherm R, Mostovoy M, Strempfer J, Habicht K, Rule K C, Dudzik E, Wolter A U B, Maljuk A and Argyriou D N 2008 Magnetic order and ferroelectricity in RMnO3 multiferroic manganites: Coupling between R- and Mn-spins *J. Phys. Condens. Matter* **20**

[9] Kimura T, Lawes G, Goto T, Tokura Y and Ramirez A P 2005 Magnetoelectric phase diagrams of orthorhombic RMnO3 (R=Gd, Tb, and Dy) *Phys. Rev. B - Condens. Matter Mater. Phys.* **71**

[10] Jirák Z, Hejtmánek J, Pollert E, Maryško M, Dlouhá M and Vratislav S 1997 Canted structures in the Mn3+/Mn4+ perovskites *J. Appl. Phys.* **81** 5790–2

[11] Muñoz A, Alonso J A, Martìnez-Lope M J, García-Muñoz J L and Fernández-Díaz M T 2000 Magnetic structure evolution of NdMnO3 derived from neutron diffraction data *J. Phys. Condens. Matter* **12** 1361–76

[12] Ribeiro J L and Vieira L G 2010 Landau model for the phase diagrams of the orthorhombic rare-earth manganites RMnO3 ( R=Eu, Gd, Tb, Dy, Ho) *Phys. Rev. B - Condens. Matter Mater. Phys.* **82**

[13] Kimura T, Goto T, Shintani H, Ishizaka K, Arima T and Tokura Y 2003 Magnetic control of ferroelectric polarization *Nature* **426** 55–8

[14] Kenzelmann M, Harris A B, Jonas S, Broholm C, Schefer J, Kim S B, Zhang C L, Cheong S W, Vajk O P and Lynn J W 2005 Magnetic inversion symmetry breaking and ferroelectricity in TbMnO3 *Phys. Rev. Lett.* **95**

[15] Agarwal H, Yadav P, Lalla N P, Alonso J A, Srivastava O N and Shaz M A 2019 Structural correlation of magneto-electric coupling in polycrystalline TbMnO3 at low temperature *J. Alloys Compd.* **806** 510–9

[16] Ratcliff W, Lynn J W, Kiryukhin V, Jain P and Fitzsimmons M R 2016 Magnetic structures and dynamics of multiferroic systems obtained with neutron scattering *npj Quantum Mater.* **1**

[17] Bousquet E and Cano A 2016 Non-collinear magnetism in multiferroic perovskites *J. Phys. Condens. Matter* **28**

[18] Sharma V, McDannald A, Staruch M, Ramprasad R and Jain M 2015 Dopant-mediated structural and magnetic properties of TbMnO3 *Appl. Phys. Lett.* **107**

[19] Dyakonov V, Bukhanko F, Kamenev V, Zubov E, Baran S, Jaworska-Goła□b T, Szytuła A, Wawrzyńska E, Penc B, Duraj R, Stüsser N, Arciszewska M, Dobrowolski W, Dyakonov K, Pientosa J, Manus O, Nabialek A, Aleshkevych P, Puzniak R, Wisniewski A, Zuberek R and Szymczak H 2006 Structural and magnetic properties of La1-x Prx Mn O3+δ (0≤x≤1.0) *Phys. Rev. B - Condens. Matter Mater. Phys.* **74**

[20] Arima T, Tokunaga A, Goto T, Kimura H, Noda Y and Tokura Y 2006 Collinear to spiral

spin transformation without changing the modulation wavelength upon ferroelectric transition in Tb1-xDyxMnO3 *Phys. Rev. Lett.* **96**

[21] Xu M H, Wang Z H, Zhang D W and Du Y W 2013 Exchange bias effect in Tb0.4Dy0.6MnO3 *J. Magn. Magn. Mater.* **340** 1–4

[22] Pavan Kumar N, Jitender T, Laxmi V and Venugopal Reddy P 2016 Thermal, magnetic and electrical properties of Tb1-xDyxMnO3 multiferroics *J. Magn. Magn. Mater.* **401** 860–9

[23] Agarwal H, Alonso J A, Muñoz Á, Choudhary R J, Srivastava O N and Shaz M A 2020 Magnetization spin reversal and neutron diffraction study of polycrystalline Tb0.55Sr0.45MnO3 *J. Alloys Compd.* **845** 156355

[24] Wu Z P, Li P G, Tang W H, Li L H and Huang Q Z 2015 Study of structure and magnetic ordering in multiferroics Tb1-xNdxMnO3 by neutron powder diffraction *J. Alloys Compd.* **644** 13–6

[25] Rodríguez-Carvajal J 1993 Recent advances in magnetic structure determination by neutron powder diffraction *Phys. B Phys. Condens. Matter* **192** 55–69

[26] Rodríguez-Carvajal J 2015 Introduction to the Program FULLPROF: Refinement of Crystal and Magnetic Structures from Powder and Single Crystal Data *Lab. Léon Brillouin (CEA-CNRS), CEA/Saclay, 91191 Gif sur Yvette Cedex,FRANCE.*

[27] Rodríguez-carvajal J 2004 BasIreps: A program for calculating irreducible representation of little groups and basis functions of polar and axial vector properties. *Solid State Phenom*

[28] Jiang N, Zhang X and Yu Y 2013 Atomic distribution, local structure and cation size effect in o-R 1-xCaxMnO3 (R = Dy, Y, and Ho) *J. Phys. Condens. Matter* **25**

[29] Iliev M N, Abrashev M V., Popov V N and Hadjiev V G 2003 Role of Jahn-Teller disorder in Raman scattering of mixed-valence manganites *Phys. Rev. B - Condens. Matter Mater. Phys.* **67**

[30] Glazer A M 1972 The classification of tilted octahedra in perovskites *Acta Crystallogr. Sect. B Struct. Crystallogr. Cryst. Chem.* **28** 3384–92

[31] Alonso J A, Martínez-Lope M J, Casais M T and Fernández-Díaz M T 2000 Evolution of the Jahn-Teller distortion of MnO6 octahedra in RMnO3 perovskites (R = Pr, Nd, Dy, Tb, Ho, Er, Y): A neutron diffraction study *Inorg. Chem.* **39** 917–23

[32] Arulraj A, Santhosh P N, Gopalan R S, Guha A, Raychaudhuri A K, Kumar N and Rao C N R 1998 Charge ordering in the rare-earth manganates: The origin of the extraordinary sensitivity to the average radius of the A-site cations, 〈rA〉 *J. Phys. Condens. Matter* **10** 8497–504

[33] Kim M W, Moon S J, Jung J H, Yu J, Parashar S, Murugavel P, Lee J H and Noh T W 2006 Effect of orbital rotation and mixing on the optical properties of orthorhombic RMnO3 (R=La, Pr, Nd, Gd, and Tb) *Phys. Rev. Lett.* **96**

[34] Kimura T, Ishihara S, Shintani H, Arima T, Takahashi T, Ishizaka K and Tokura Y 2003 Distorted perovskite with eg1 configuration as a frustrated spin system *Phys. Rev. B - Condens. Matter Mater. Phys.* **68**

[35] Kováčik R, Murthy S S, Quiroga C E, Ederer C and Franchini C 2016 Combined first-principles and model Hamiltonian study of the perovskite series RMnO3 (R=La,Pr,Nd,Sm,Eu, and Gd) *Phys. Rev. B* **93**

[36] RADO G T and SUHL H 1963 Preface *Magnetism: Spin Arrangements and Crystal Structure, Domains, and Micromagnetics* ed G T RADO and H SUHL (Academic Press) p VOL III

**Table Captions:**

**Table 1:** Structural Parameters for $TbMnO_3$, $Tb_{0.6}Pr_{0.4}MnO_3$ and $PrMnO_3$ at room temperature

**Table 2:** Geometrical parameters, interatomic distances (Å) and angles (º) characterizing the crystal structure of TMO, TPMO and PMO at room temperature observation of XRD

**Table 3:** *dc* Magnetization analysis of polycrystalline $Tb_{0.6}Pr_{0.4}MnO_3$ by Curie-Weiss fitting

**Table 4:** Neutron powder diffraction refinement parameters of polycrystalline $Tb_{0.6}Pr_{0.4}MnO_3$ at various temperatures down to 1.5 K

**Table 5:** Refined structural variation in polycrystalline $Tb_{0.6}Pr_{0.4}MnO_3$ observed form Rietveld refinement of Neutron powder diffraction patterns

**Table 6:** Notation for Mn and Tb ion according to Group theory for space group Pnma for q = (0,0,0). The notation for the positions of the Mn atoms (site 4b ) is: 1 (0,0,1/2); 2 (1/2,0,0); 3 (0,1/2,1/2); 4 (1/2,1/2,0). For the Tb atoms (site 4c ) the notation is: 1 (x,y,z), 2 (-x+1/2,-y,z+1/2), 3 (-x,y+1/2,-z) and 4 (x+1/2,-y+1/2,-z+1/2).

**Table 7:** Magnetic moment ($\mu_B$) of different cations at various temperatures for $Tb_{0.6}Pr_{0.4}MnO_3$

**Figure Captions:**

**Figure 1**: (a) Comparison of X-ray diffraction patterns of $TbMnO_3$ (TMO- shown in red), $Tb_{0.6}Pr_{0.4}MnO_3$ (TPMO- shown in Blue), $PrMnO_3$ (PMO- shown in Violet); the inset is showing the shifting of XRD peaks to higher symmetry order (b) Rietveld refinement of $Tb_{0.6}Pr_{0.4}MnO_3$ (c-e) structural view of TMO, TPMO and PMO, respectively, having $a^-b^+a^-$ type octahedral distortion with orthorhombic *Pnma* symmetry

**Figure 2**: (a) ZFC-FC DC magnetization of TPMO in the temperature range of 5 K – 300 K and inset is showing the clear bifurcation in ZFC and FC curve (b) first derivative of magnetization with respect to temperature giving the phase transition temperature (c) inverse susceptibility showing the AFM nature of $Tb_{0.6}Pr_{0.4}MnO_3$ and inset is showing different magnetic ordering tending from AFM to FM ordering (d) difference in between FC and ZFC curve of magnetization

**Figure 3:** Comparison among the *dc* magnetization and inverse susceptibility for polycrystalline $Tb_{0.6}Pr_{0.4}MnO_3$ at 100 Oe, 5000 Oe, and 10 kOe.

**Figure 4**: Field-dependent dc magnetization (M/H) studies for $Tb_{0.6}Pr_{0.4}MnO_3$ up to 70 kOe at different temperatures; the inset is showing the weak coercivity at 30 K and 5 K.

**Figure 5** (a) Comparison of Neutron powder diffraction patterns of $Tb_{0.6}Pr_{0.4}MnO_3$ at various temperatures, where the inset is showing the magnetic peaks arising at a lower angle (b) difference curve of NPD pattern showing the magnetic reflections at 50 K, 25 K and 1.5 K.

**Figure 6** Rietveld refinements of NPD patterns along with the nuclear structures at (a) 300 K, (b) 150 K, (c) 50 K, and (d) 25 K. The inset of (c) and (d) is showing the magnetic contribution due to Mn ordering.

**Figure 6** Rietveld refinements of NPD patterns along with the nuclear structures at (e) 1.5 K showing the magnetic contribution due to Mn and Tb ordering

**Figure 7:** Magnetic structures at (a) 50K, (b) 25 and (c) 1.5 K for the $Tb_{0.6}Pr_{0.4}MnO_3$ representing A-type antiferromagnetic ordering. The Tb moments appear ordered only at 1.5 K, adopting a ferromagnetic arrangement along the negative b-direction.

**Figure 8:** Phase diagram representing the magnetic phase transition in $TbMnO_3$, $Tb_{0.6}Pr_{0.4}MnO_3$, and $PrMnO_3$ along with the Mn-O-Mn bond angle and JT distortion variation

**Table 1:**

**Structural Parameters for $TbMnO_3$, $Tb_{0.6}Pr_{0.4}MnO_3$ and $PrMnO_3$ at room temperature**

| Composition | Lattice Parameter | Atoms | Atomic Positions x | y | z | Wyckoff position | Occupancy |
|---|---|---|---|---|---|---|---|
| **$TbMnO_3$ (*Pnma*)** | a = 5.839(9) Å, b = 7.399(8) Å, c = 5.297(8) Å V = 228.939(6) Å$^3$ | **Tb** | 0.9190(2) | 0.25 | 0.0161(7) | 4c | 1 |
| | | **Mn** | 0 | 0 | 0.5 | 4b | 1 |
| | | **O1** | 0.0292(6) | 0.25 | 0.6043(2) | 4c | 1 |
| | | **O2** | 0.6721(1) | 0.4500(1) | 0.2970(1) | 8d | 1 |
| **$Tb_{0.6}Pr_{0.4}MnO_3$ (*Pnma*)** | a = 5.811(4) Å, b = 7.410(1) Å, c = 5.348(6) Å V = 232.193(5) Å$^3$ | **Tb** | 0.9245(9) | 0.25 | 0.0159(5) | 4c | 0.6 |
| | | **Pr** | 0.9245(9) | 0.25 | 0.0159(5) | 4c | 0.4 |
| | | **Mn** | 0 | 0 | 0.5 | 4b | 1 |
| | | **O1** | 0.0220(6) | 0.25 | 0.5912(3) | 4c | 1 |
| | | **O2** | 0.6879(7) | 0.0452(4) | 0.2902(6) | 8d | 1 |
| **$PrMnO_3$ (*Pnma*)** | a = 5.573(9) Å, b = 7.674(9) Å, c = 5.458(1) Å V = 233.488(4) Å$^3$ | **Pr** | 0.9582(7) | 0.25 | 0.0092(1) | 4c | 1 |
| | | **Mn** | 0 | 0 | 0.5 | 4b | 1 |
| | | **O1** | 0.0170(2) | 0.25 | 0.5891(2) | 4c | 1 |
| | | **O2** | 0.7197(7) | 0.03439 | 0.2543(1) | 8d | 1 |

**Table 2:**

**Geometrical parameters, interatomic distances (Å) and angles (º) characterizing the crystal structure of TMO, TPMO and PMO at room temperature observation of XRD**

| Magnetic Properties | TMO | TPMO | PMO |
|---|---|---|---|
| **Mn-O1** | 1.983(8) | 1.935(2) | 1.982(4) |
| **Mn-O2** | 1.903(4) | 1.928(2) | 1.869(1) |
| **Mn-O2´** | 2.227(3) | 2.158(6) | 2.075(2) |
| **<Mn-O>** | 2.022(7) | 2.007(5) | 1.975(3) |
| **Mn-O1-Mn** | 145.27(1) | 150.08(7) | 151.03(4) |
| **Mn-O2-Mn** | 145.17(5) | 149.77(3) | 162.67(5) |
| **<Mn-O-Mn>** | 145.22(3) | 149.93(5) | 156.85(4) |
| **<w> = pi - <Mn-O-Mn>** | 34.77(7) | 15.03(5) | 23.145(5) |
| **One electron bandwidth ‘W'** | 0.081(2) eV | 0.084(5) eV | 0.090(4) eV |
| **Jahn teller distortion $\Delta_{JT}$** | 0.0051(6) | 0.0028(2) | 0.0018(1) |
| **<$r_A$>** | 1.095(4) | 1.128(6) | 1.179(1) |
| **Strain Parameter ‘s’** | 0.097 | 0.082 | 0.0104 |

**Table 3:**

***dc* Magnetization analysis of polycrystalline $Tb_{0.6}Pr_{0.4}MnO_3$ by Curie-Weiss fitting**

| Magnetic Properties | 100 Oe | 5000 Oe | 10 kOe |
|---|---|---|---|
| **Curie Constant** | 11.71(5) | 13.05(6) | 13.09(5) |
| $\mu_{eff}$ | 9.67(6) $\mu_B$ | 10.21(5) $\mu_B$ | 10.23(1) $\mu_B$ |
| **Curie Temperature** | -24 K | -25 K | -26 K |

**Table 4:**

**Neutron powder diffraction refinement parameters of polycrystalline $Tb_{0.6}Pr_{0.4}MnO_3$ at various temperatures down to 1.5 K**

| Parameters | 300 K | 150 K | 50 K | 25 K | 1.5 K |
|---|---|---|---|---|---|
| **a (Å)** | 5.8012(7) | 5.7962(4) | 5.7833(5) | 5.7799(5) | 5.7806(2) |
| **b (Å)** | 7.4572(9) | 7.4451(2) | 7.4413(4) | 7.4385(3) | 7.4399(1) |
| **c (Å)** | 5.3387(1) | 5.3375(9) | 5.3453(1) | 5.3461(5) | 5.3473(1) |
| **b/√2 (Å)** | 5.2730(3) | 5.2644(6) | 5.2617(5) | 5.2598(1) | 5.2608(1) |
| **Mn-O1/Mn-O4 (Å)** | 1.9413(1) | 1.9397(3) | 1.9392(4) | 1.9386(1) | 1.9391(2) |
| **Mn-O2/Mn-O5 (Å)** | 1.9042(6) | 1.9036(8) | 1.9078(9) | 1.9064(0) | 1.9078(2) |
| **Mn-O3/Mn-O6 (Å)** | 2.2032(1) | 2.2008(4) | 2.1934(7) | 2.1947(6) | 2.1921(9) |
| **Mn-O1-Mn (°)** | 147.621(6) | 147.298(5) | 147.199(3) | 147.179(3) | 147.147(4) |
| **Mn-O2-Mn (°)** | 147.271(4) | 147.193(0) | 147.429(4) | 147.350(9) | 147.528(1) |
| **Volume ($Å^3$)** | 230.96(2) | 230.18(5) | 230.03(9) | 229.85(4) | 229.97(2) |
| **Bragg R-factor (%)** | 2.564 | 2.492 | 2.671 | 2.678 | 3.108 |
| **$R_f$ factor (%)** | 1.579 | 1.558 | 1.658 | 1.715 | 1.804 |
| **Chi Square** | 1.40 | 1.70 | 1.88 | 2.58 | 2.75 |
| **Magnetic R factor (%)** | - | - | 18.1 | 14.5 | 9.5 |

**Table 5:**

**Refined structural variation in polycrystalline $Tb_{0.6}Pr_{0.4}MnO_3$ observed form Rietveld refinement of Neutron powder diffraction patterns. In all cases, the space group is *Pnma*. U parameters are given in Å$^2$.**

| | | 300 K | 150 K | 50 K | 25 K | 1.5 K | Site |
|---|---|---|---|---|---|---|---|
| Tb | x | 0.9233(8) | 0.9241(1) | 0.9245(1) | 0.9258(7) | 0.9235(8) | 4c |
| | y | 0.25 | 0.25 | 0.25 | 0.25 | 0.25 | |
| | z | 0.0146(4) | 0.0135(1) | 0.0133(2) | 0.0129(8) | 0.0127(6) | |
| | U | 0.01 | 0.006 | 0.004 | 0.004 | 0.003 | |
| Pr | x | 0.9233(8) | 0.9241(1) | 0.9245(1) | 0.9258(7) | 0.9235(8) | 4c |
| | y | 0.25 | 0.25 | 0.25 | 0.25 | 0.25 | |
| | z | 0.0146(4) | 0.0135(1) | 0.0133(2) | 0.0129(8) | 0.0127(6) | |
| | U | 0.01 | 0.006 | 0.004 | 0.004 | 0.003 | |
| Mn | x | 0 | 0 | 0 | 0 | 0 | 4b |
| | y | 0 | 0 | 0 | 0 | 0 | |
| | z | 0.5 | 0.5 | 0.5 | 0.5 | 0.5 | |
| | U | 0.006 | 0.003 | 0.002 | 0.002 | 0.002 | |
| O1 | x | 0.0294(5) | 0.0300(2) | 0.0296(8) | 0.0291(4) | 0.0292(5) | 4c |
| | y | 0.25 | 0.25 | 0.25 | 0.25 | 0.25 | |
| | z | 0.596(2) | 0.5969(8) | 0.5972(7) | 0.5974(8) | 0.5975(5) | |
| | U | 0.01 | 0.006 | 0.005 | 0.004 | 0.004 | |
| O2 | x | 0.6775(3) | 0.6776(7) | 0.6790(6) | 0.6787(2) | 0.6791(7) | 8d |
| | y | 0.0478(2) | 0.0483(4) | 0.0485(7) | 0.0487(2) | 0.0482(9) | |
| | z | 0.2924(9) | 0.2923(7) | 0.2920(5) | 0.2919(4) | 0.2920(1) | |
| | U | 0.011 | 0.007 | 0.004 | 0.005 | 0.004 | |

**Table 6:**

**Notation for Mn and Tb ion according to Group theory for space group *Pnma* for q = (0,0,0). The notation for the positions of the Mn atoms (site 4b) is: 1 (0,0,1/2); 2 (1/2,0,0); 3 (0,1/2,1/2); 4 (1/2,1/2,0). For the Tb atoms (site 4c) the notation is: 1 (x,y,z), 2 (-x+1/2,-y,z+1/2), 3 (-x,y+1/2,-z) and 4 (x+1/2,-y+1/2,-z+1/2).**

| | Mn | | Tb | | Space Group |
|---|---|---|---|---|---|
| $\Gamma_1$ | ($A_x$, $G_y$, $C_z$) | Mn_1 (u, v, w)<br>Mn_2 (–u, -v, w)<br>Mn_3 (–u, v, -w)<br>Mn_4 (u, -v, -w) | (0, $G_y$, 0) | Tb_1 (0, u, 0)<br>Tb_2 (0, -u, 0)<br>Tb_3 (0, u, 0)<br>Tb_4 (0, -u, 0) | Pnma |
| $\Gamma_2$ | | | ($A_x$, 0, $C_z$) | Tb_1 (u, 0, v)<br>Tb_2 (-u, 0, v)<br>Tb_3 (-u, 0, -v)<br>Tb_4 (u, 0, -v) | Pn´m´a´ |
| $\Gamma_3$ | ($G_x$, $A_y$, $F_z$) | Mn_1 (u, v, w)<br>Mn_2 (–u, -v, w)<br>Mn_3 (u, -v, w)<br>Mn_4 (-u, v, w) | ($G_x$, 0, $F_z$) | Tb_1 (u, 0, v)<br>Tb_2 (-u, 0, v)<br>Tb_3 (u, 0, v)<br>Tb_4 (-u, 0, v) | Pnm´a´ |
| $\Gamma_4$ | | | (0, $A_y$, 0) | Tb_1 (0, u, 0)<br>Tb_2 (0, -u, 0)<br>Tb_3 (0, -u, 0)<br>Tb_4 (0, u, 0) | Pnm´a |
| $\Gamma_5$ | ($C_x$, $F_y$, $A_z$) | Mn_1 (u, v, w)<br>Mn_2 (u, v, -w)<br>Mn_3 (–u, v,-w)<br>Mn_4 (-u, v, w) | (0, $F_y$, 0) | Tb_1 (0, u, 0)<br>Tb_2 (0, u, 0)<br>Tb_3 (0, u, 0)<br>Tb_4 (0, u, 0) | Pn´ma´ |
| $\Gamma_6$ | | | ($C_x$, 0, $A_z$) | Tb_1 (u, 0, v)<br>Tb_2 (u, 0, -v)<br>Tb_3 (-u, 0, -v)<br>Tb_4 (-u, 0, v) | Pn´ma |
| $\Gamma_7$ | ($F_x$, $C_y$, $G_z$) | Mn_1 (u, v, w)<br>Mn_2 (u, v, -w)<br>Mn_3 (u, -v, w)<br>Mn_4 (u, -v, -w) | ($F_x$, 0, $G_z$) | Tb_1 (u, 0, v)<br>Tb_2 (u, 0, -v)<br>Tb_3 (u, 0, v)<br>Tb_4 (u, 0, -v) | Pn´m´a |
| $\Gamma_8$ | | | (0, $C_y$, 0) | Tb_1 (0, u, 0)<br>Tb_2 (0, u, 0)<br>Tb_3 (0, -u, 0)<br>Tb_4 (0, -u, 0) | Pnma´ |

**Table 7:**

**Magnetic moment ($\mu_B$) of different cations at various temperatures for $Tb_{0.6}Pr_{0.4}MnO_3$**

| Temperature | Cation | $m_x$ | $m_y$ | $m_z$ | Magnetic Moment |
|---|---|---|---|---|---|
| **50 K** | $Mn^{3+}$ | 1.19(7) | - | - | 1.19(7) |
| **25 K** | $Mn^{3+}$ | 2.32(2) | - | - | 2.32(2) |
| **1.5 K** | $Mn^{3+}$ | 2.57(3) | - | - | 2.57(3) |
| | $Tb^{3+}/Pr^{3+}$ | 0 | -0.57(2) | 0 | 0.57(2) |

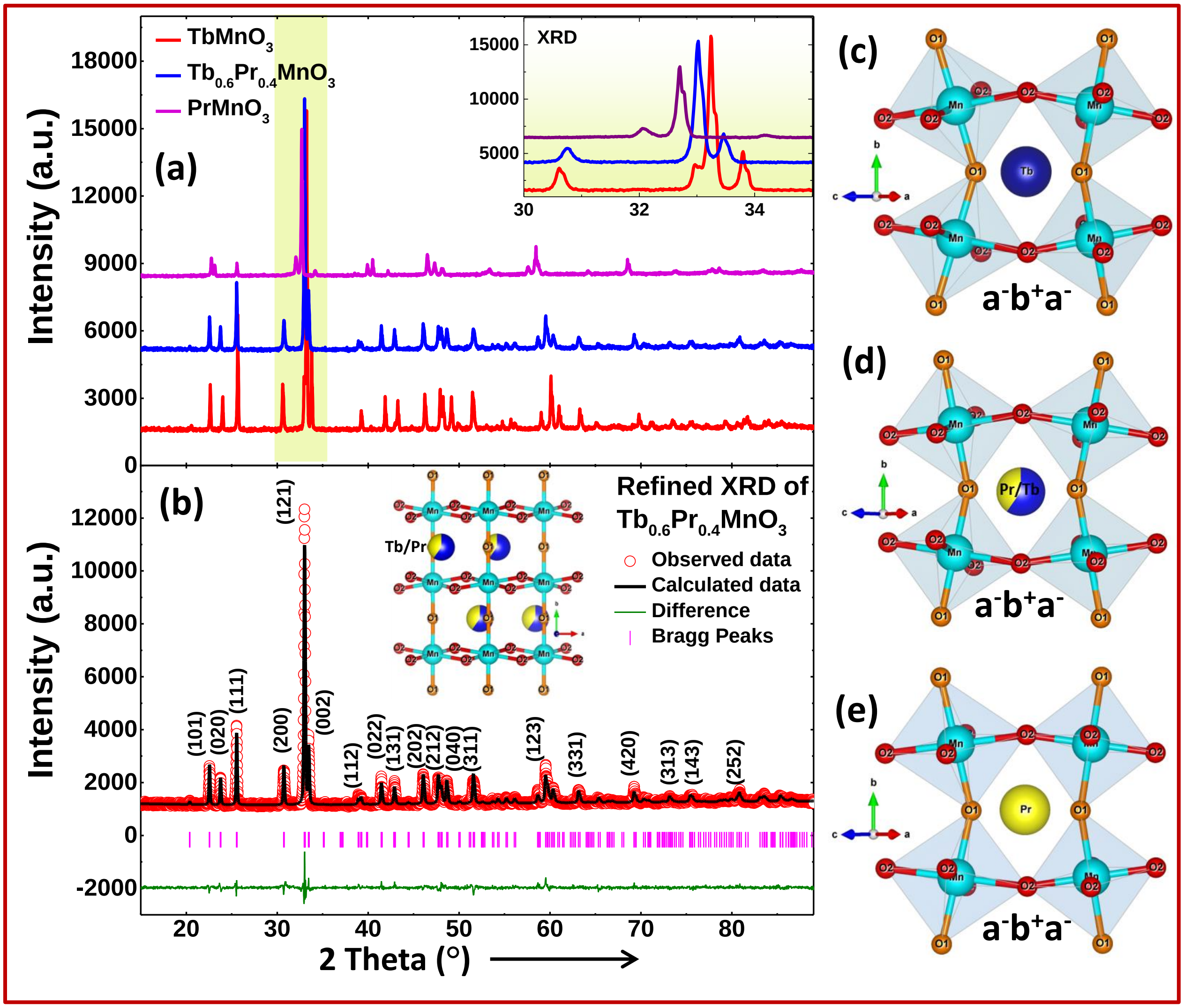

**Figure 1**: (a) Comparison of X-ray diffraction patterns of $TbMnO_3$ (TMO- shown in red), $Tb_{0.6}Pr_{0.4}MnO_3$ (TPMO- shown in Blue), $PrMnO_3$ (PMO- shown in Violet); the inset is showing the shifting of XRD peaks to higher symmetry order (b) Rietveld refinement of $Tb_{0.6}Pr_{0.4}MnO_3$ (c-e) structural view of TMO, TPMO and PMO, respectively, having $a^-b^+a^-$ type octahedral distortion with orthorhombic *Pnma* symmetry

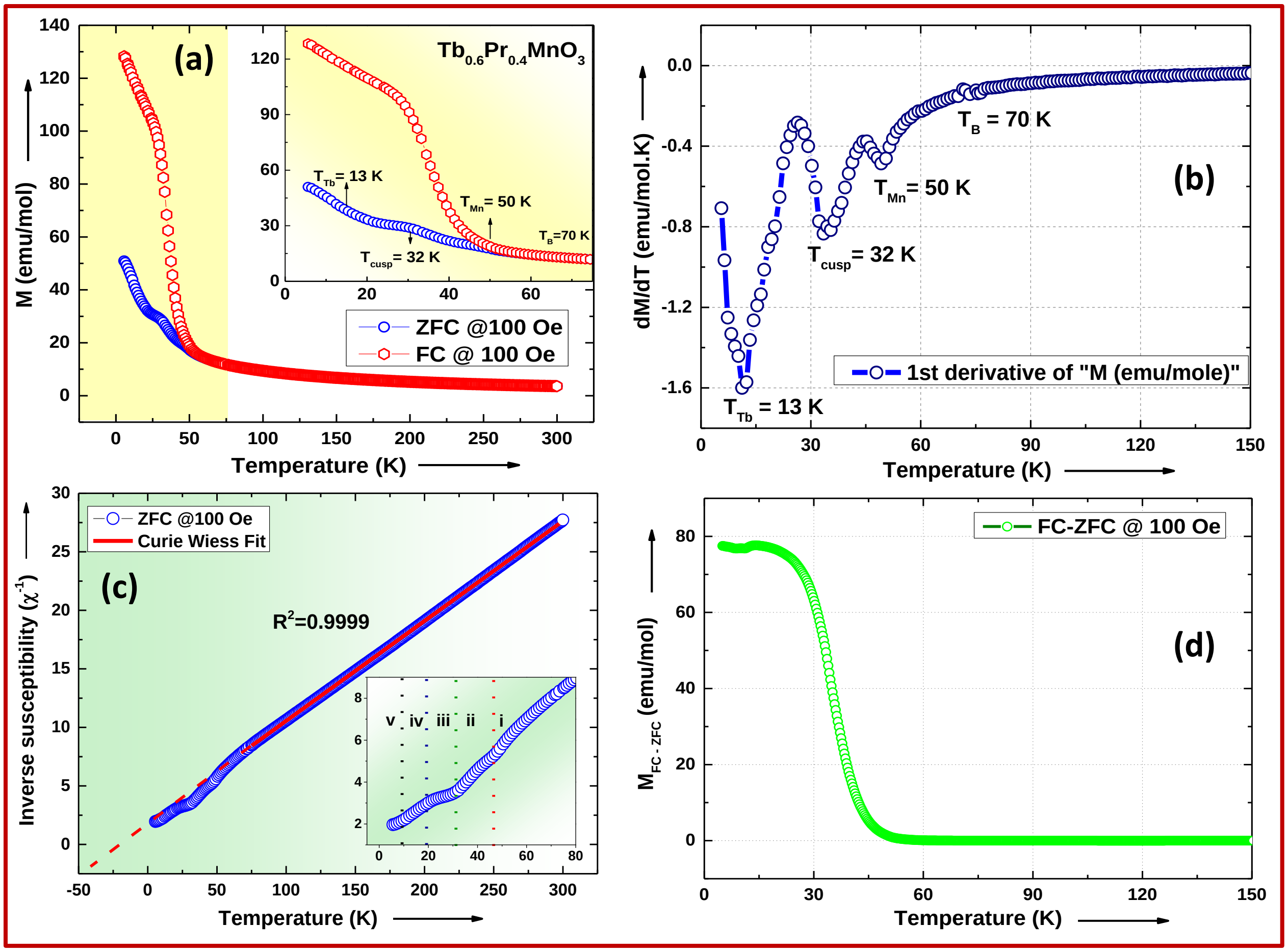

**Figure 2**: (a) Zero Field Cooled (ZFC)-Field cooled (FC) DC magnetization of TPMO in the temperature range of 5 K – 300 K and inset is showing the clear bifurcation in ZFC and FC curve (b) first derivative of magnetization with respect to temperature giving the phase transition temperature (c) inverse susceptibility showing the AFM nature of $Tb_{0.6}Pr_{0.4}MnO_3$ and inset is showing different magnetic ordering tending from AFM to FM ordering (d) difference in between FC and ZFC curve of magnetization

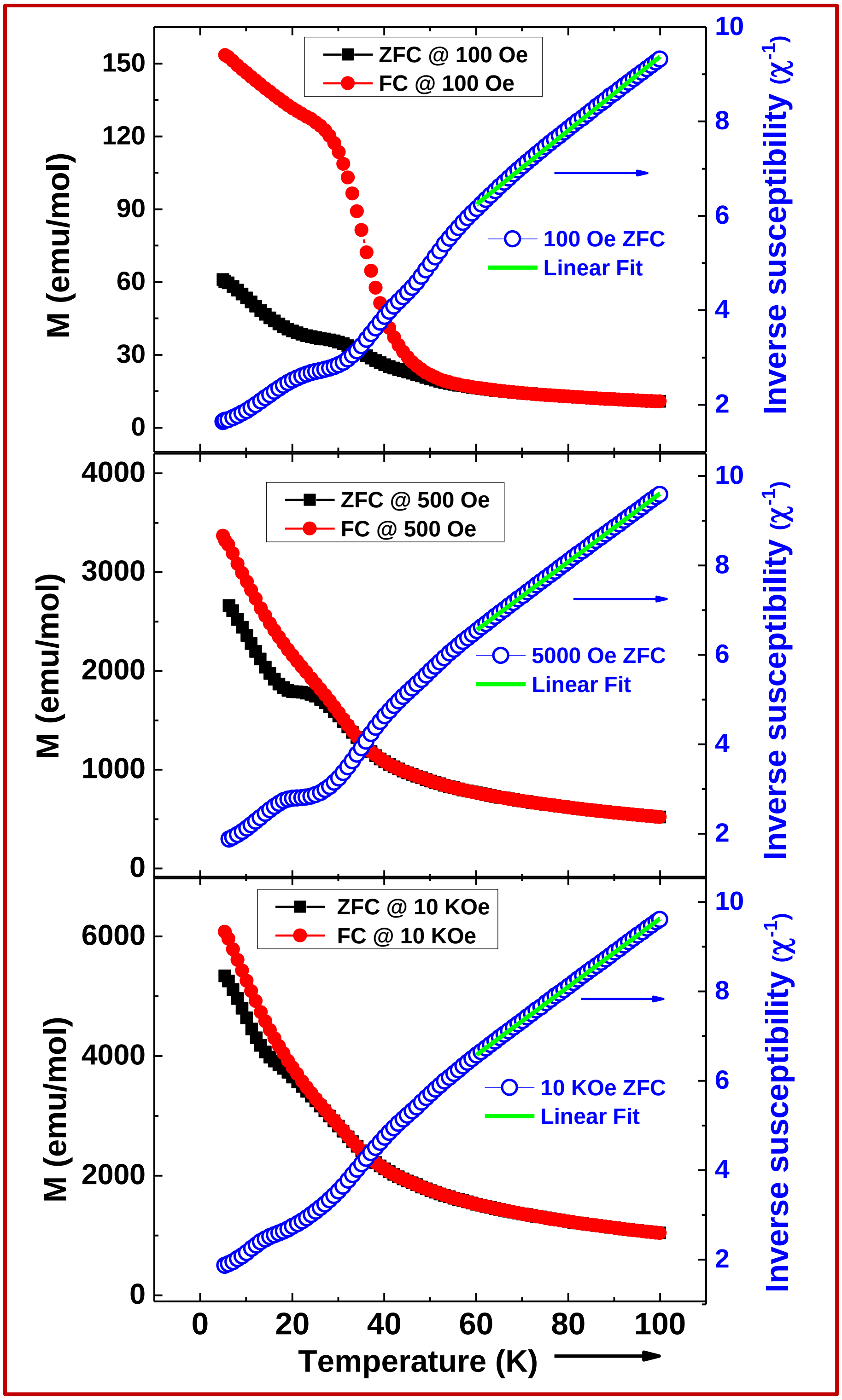

**Figure 3:** Comparison among the *dc* magnetization and inverse susceptibility for polycrystalline $Tb_{0.6}Pr_{0.4}MnO_3$ at 100 Oe, 5000 Oe, and 10 kOe.

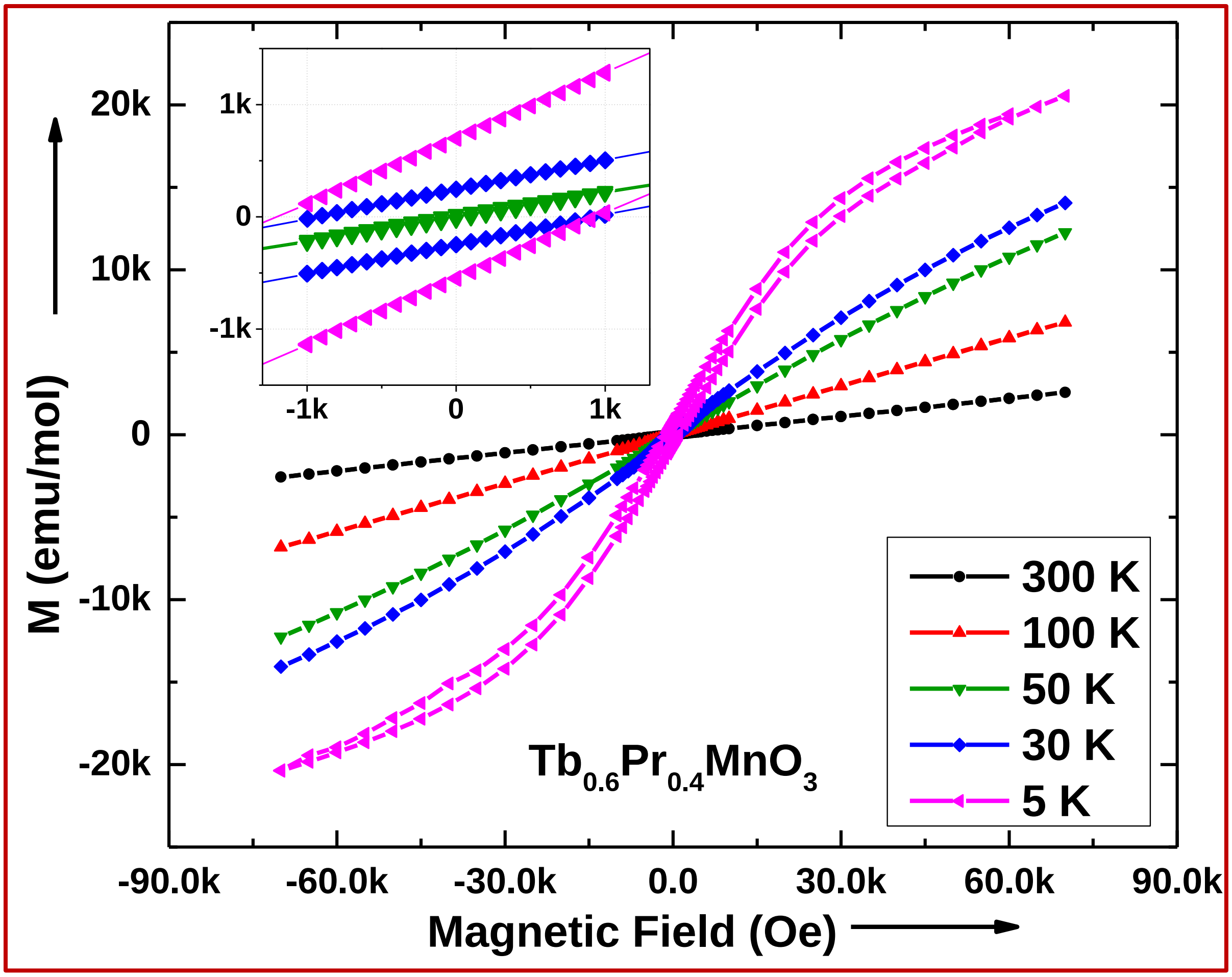

**Figure 4**: Field-dependent dc magnetization (M/H) studies for $Tb_{0.6}Pr_{0.4}MnO_3$ up to 70 kOe at different temperatures; the inset is showing the weak coercivity at 30 K and 5 K.

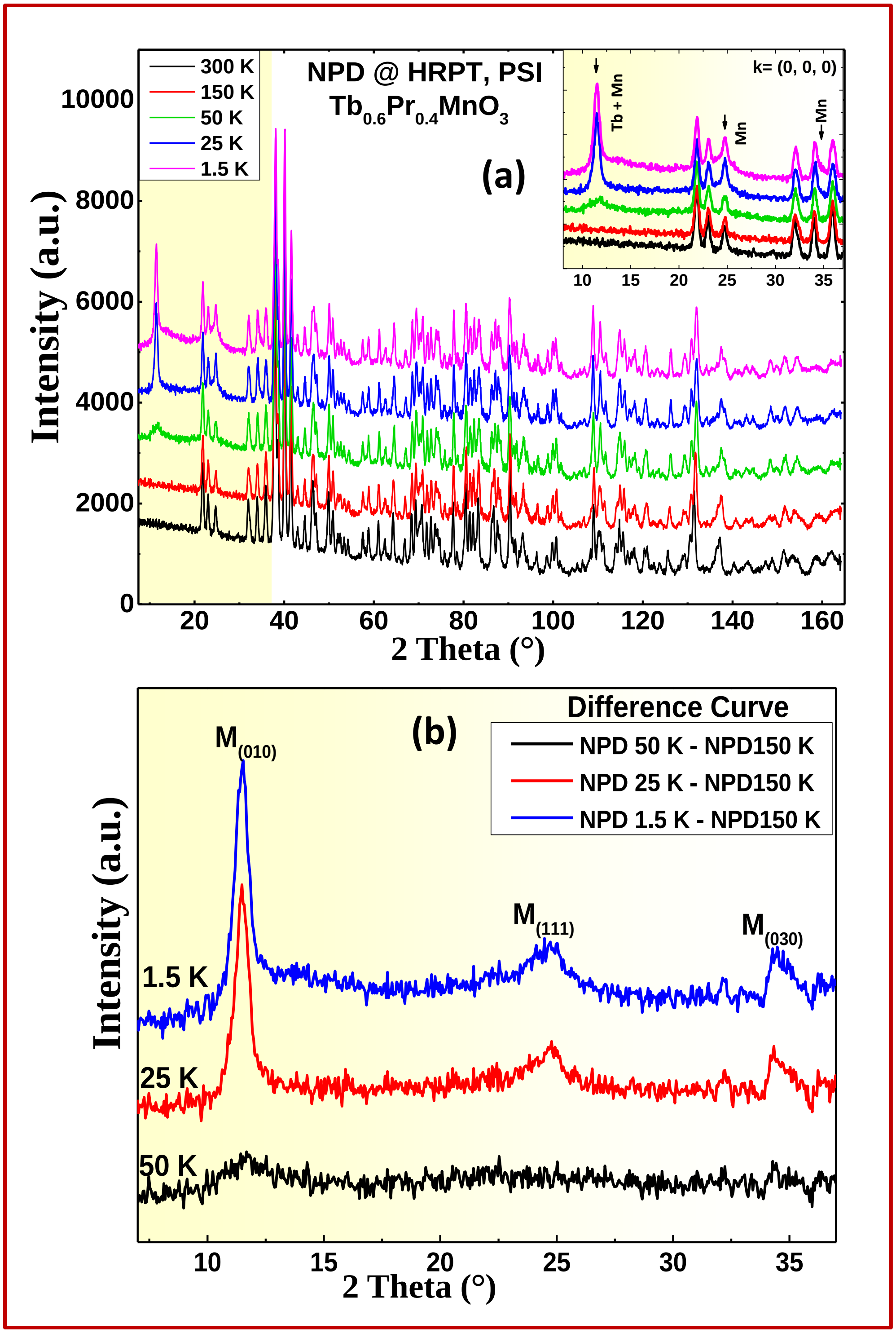

**Figure 5** (a) Comparison of Neutron powder diffraction patterns of $Tb_{0.6}Pr_{0.4}MnO_3$ at various temperatures, where the inset is showing the magnetic peaks arising at a lower angle (b) difference curve of NPD pattern showing the magnetic reflections at 50 K, 25 K and 1.5 K.

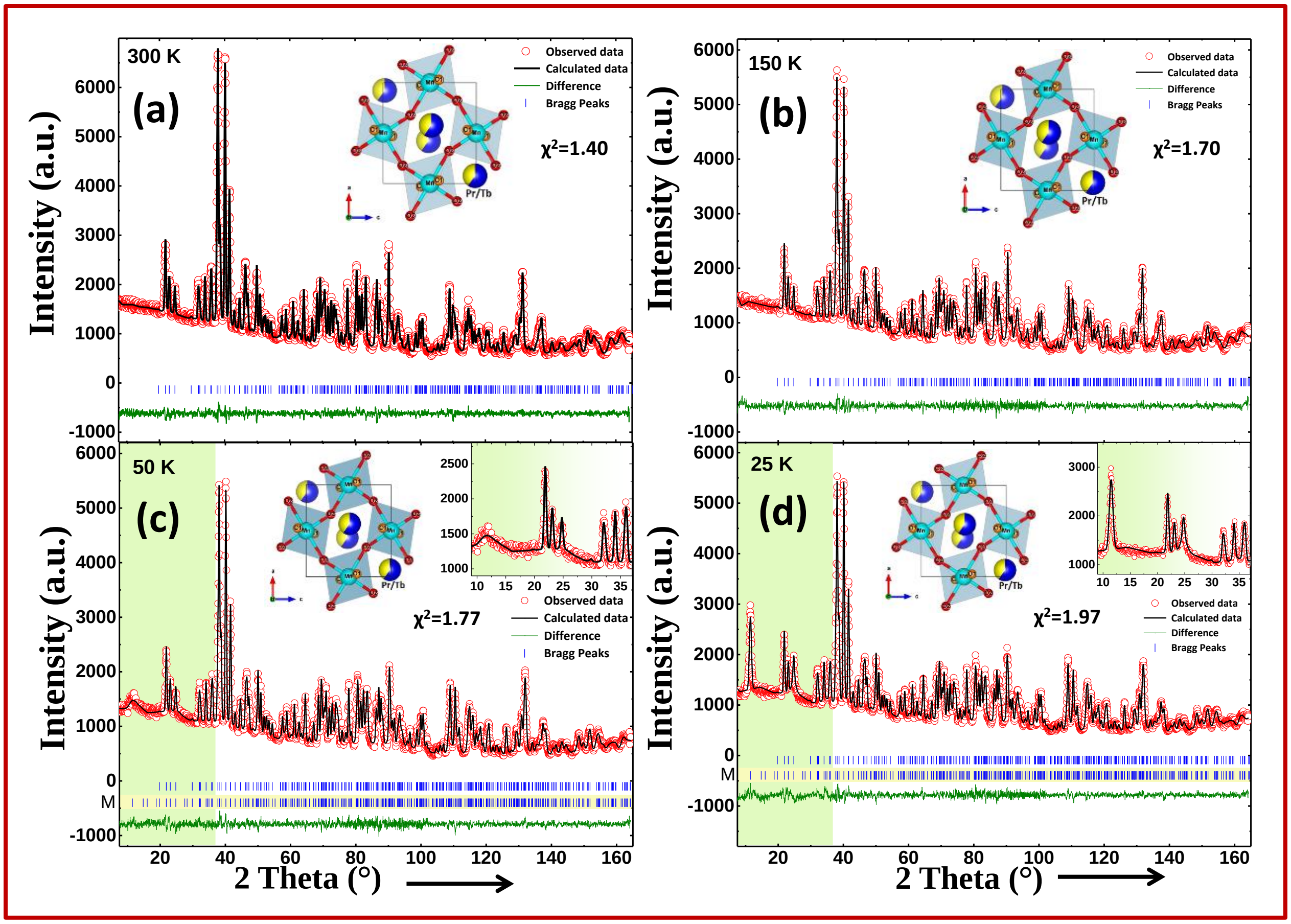

**Figure 6** Rietveld refinements of NPD patterns along with the nuclear structures at (a) 300 K, (b) 150 K, (c) 50 K, and (d) 25 K. The inset of (c) and (d) is showing the magnetic contribution due to Mn ordering.

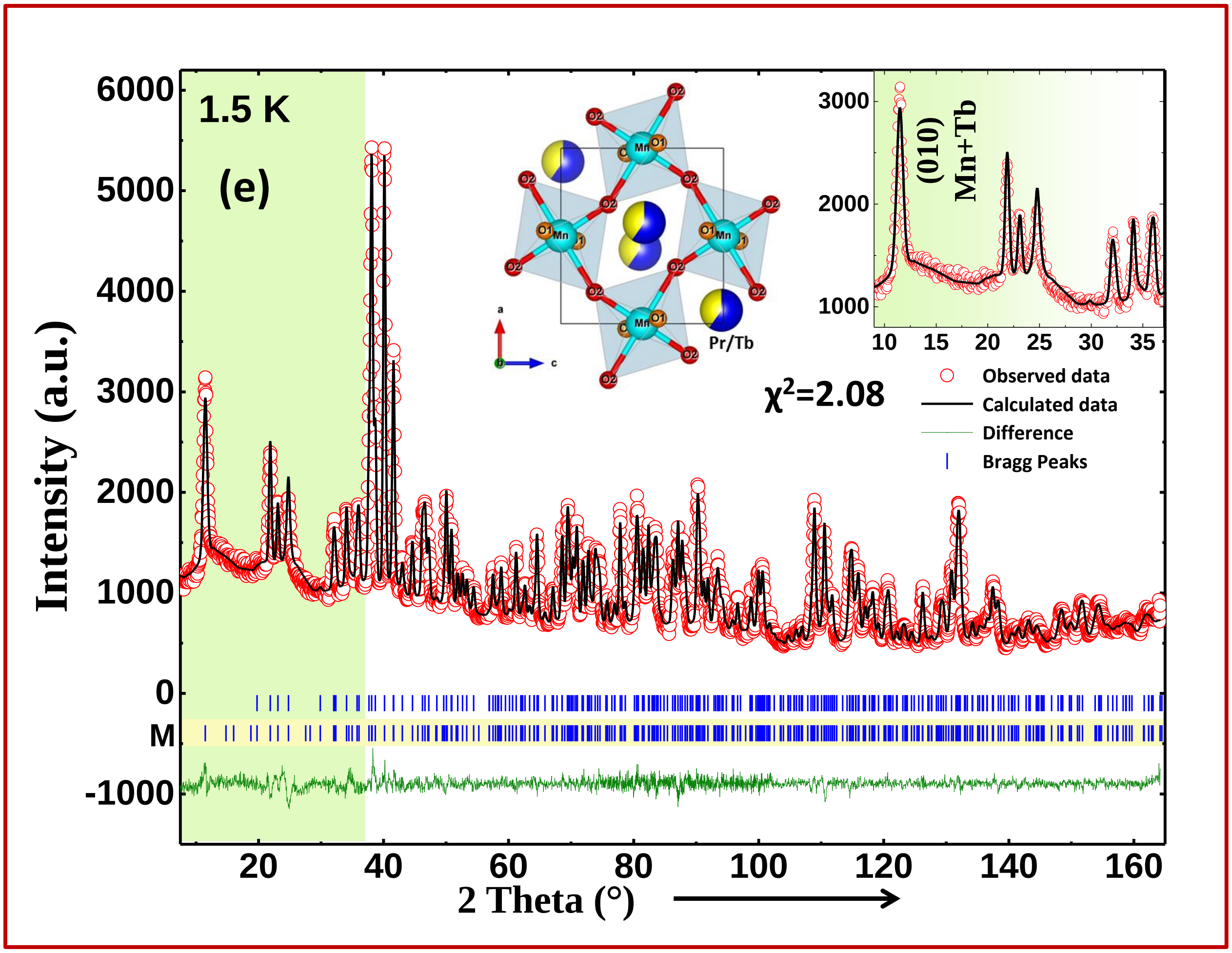

**Figure 6** Rietveld refinements of NPD patterns along with the nuclear structures at (e) 1.5 K showing the magnetic contribution due to Mn and Tb ordering

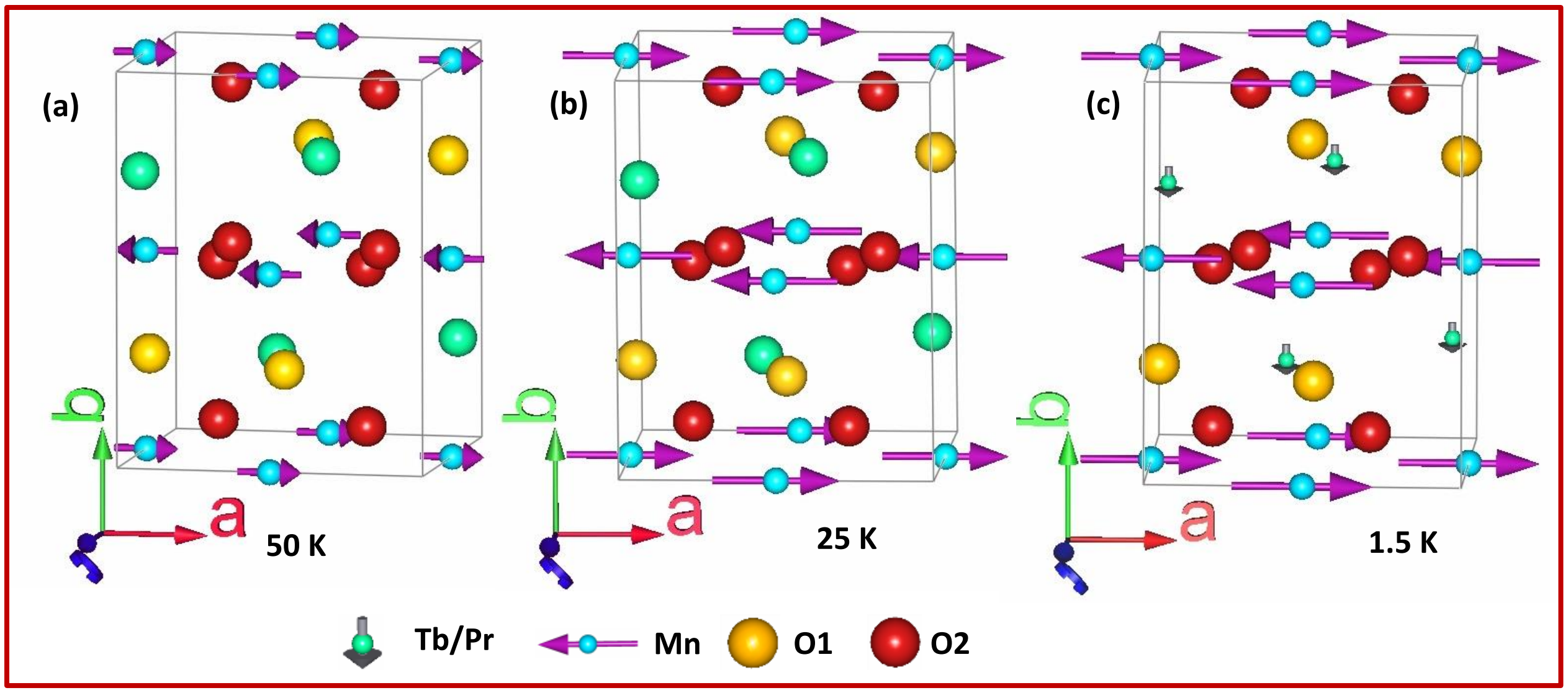

**Figure 7:** Magnetic structures at (a) 50K, (b) 25 and (c) 1.5 K for $Tb_{0.6}Pr_{0.4}MnO_3$ representing A-type antiferromagnetic ordering. The Tb moments appear ordered only at 1.5 K, adopting a ferromagnetic arrangement along the negative b-direction.

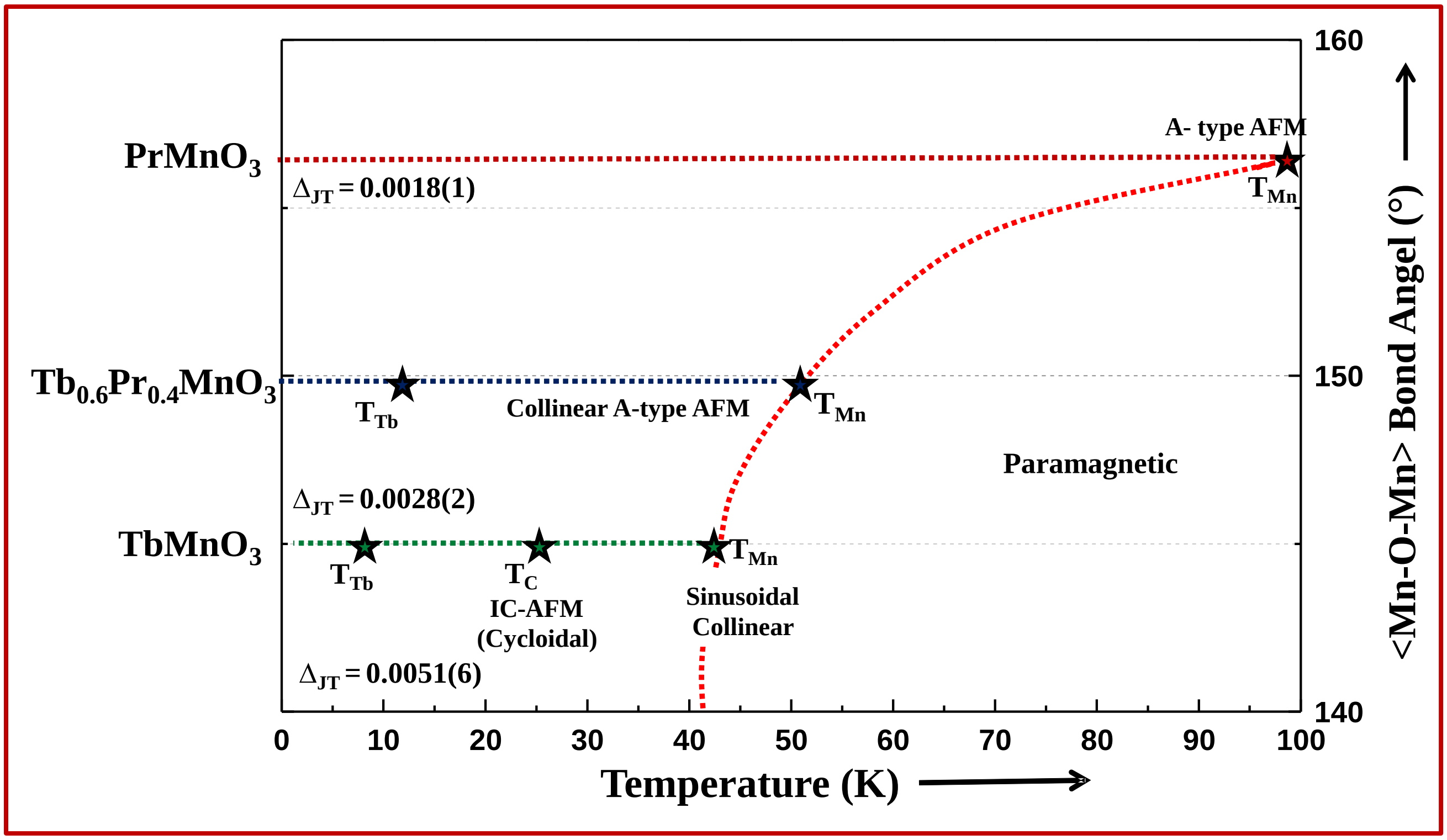

**Figure 8:** Phase diagram representing the magnetic phase transition in $TbMnO_3$, $Tb_{0.6}Pr_{0.4}MnO_3$, and $PrMnO_3$ along with the Mn-O-Mn bond angle and JT distortion variation